%% file: main.tex
\documentclass[
]{ceurart}

\usepackage{listings}
\begin{document}
\copyrightyear{2026}
\copyrightclause{Copyright for this paper by its authors.
  Use permitted under Creative Commons License Attribution 4.0
  International (CC BY 4.0).}

\conference{Beyond Algorithms: A Workshop on the Interdisciplinarity of Recommender Systems (BEYOND 2026), October 2nd, 2026, co-located with the 20th ACM Recommender Systems Conference, Minneapolis, Minnesota, USA.}

\title{A Pedagogically Demonstrative Model Visualizing the Pathway from Online Interactions to Personalized Recommendation}






\author[1]{Sushmita Khan}[%
email=sushmik@clemson.edu,
]

\author[2]{Connor Pennington}[%
email=cjpenni@clemson.edu,
]

\author[3]{Bart P. Knijnenburg}[%
email=bartk@clemson.edu,
]

\address[]{Clemson University, Clemson, SC}

\begin{abstract}
Personal digital activity increasingly shapes online experiences, yet few users have been educated regarding the processes transforming raw interactions into personalized suggestions. We developed an education artifact that illustratively simulates how AI leverages users' digital activities to shape online recommendations (e.g., ads). Our artifact processes users' digital activity using a locally-hosted LLM to generate user profiles of their inferred interests and personalized recommendations. A three-layered Sankey diagram maps data sources through inferred interests to personalized recommendations. Interactive filters enable users to explore how different combinations of data sources influence personalized outcomes. This paper describes the artifact and its educational value, and reports findings of a pilot think-aloud study with six young adults. We find that the artifact effectively taught participants the conceptual relationship between digital activities and personalized recommendations. While this did lead participants to develop privacy awareness, they anticipated minimal behavior change due to the perceived unavoidability of platform participation.
\end{abstract}


\maketitle

\input{01_introduction}
\input{02_visualizationDesign}
\input{03_thinkAloudStudy}
\input{04_discussionConclusion}

\bibliography{bibliography.bib}

\end{document}

%% file: 01_introduction.tex
\section{Introduction and Related Work}
 
As digital platforms increasingly integrate machine learning and behavior modeling, personalization affects an ever-growing number of users~\cite{10.1145/3708319.3733690}. Most users have not been taught how their digital activities shape the personalization they encounter~\cite{ghori2022theory}, leading to frequent misconceptions~\cite{iqbalTeachingAIawareness2025, foley_is_2019}. This lack of understanding is problematic, particularly for younger adults, as it increases their vulnerability to manipulation, discrimination, and a lack of autonomy when navigating online spaces~\cite{10.1145/3613904.3642460, zhao2019sillyname, khan2025inclusive}. Extant literature reports that younger adults comprehend algorithmic decision-making best when presented tangibly, for example, through visual representation~\cite{livingstone2019audiences, singh2022trust}. As such, to facilitate young adults' understanding of how their digital activities shape their online experiences, we developed an educational artifact that illustratively simulates the complex relationship between users' personal digital activity, user profiling, and personalized recommendations.

Our educational artifact takes a user's digital activities from a number of data sources (e.g. Google Takeout) as input, feeds them to a locally hosted Large Language Model (LLM) to infer the user's interests and a set of personalized recommendations, and then generates a three-layered Sankey diagram~\cite{riehmann2005interactive} to visualize the flow between data sources, inferred interests, and recommendations (see Figure~\ref{fig:prototype} and Section~\ref{sec:viz_design} for details). Following the development of this artifact, we conducted a pilot think-aloud study with six young adults (Section~\ref{sec:results}) to answer the following research question: \emph{To what extent does our Sankey diagram help users conceptualize the relationship between digital activities and personalized} recommendations (e.g. online ads)? Results of the pilot study show that simulating and visualizing the user profiling and personalization process indeed helps users understand how their online activities shape their online experiences. Furthermore, we find that this increased understanding leads participants to develop their privacy awareness. Nevertheless, participants noted that they did not anticipate changing their behavior, as behavioral tracking is unavoidable on most online platforms, and most users find themselves unwilling to leave these platforms.

%% file: 02_visualizationDesign.tex
\section{Educational Artifact Development} \label{sec:viz_design}

\begin{figure}
    \centering
    \includegraphics[width=0.65\linewidth]{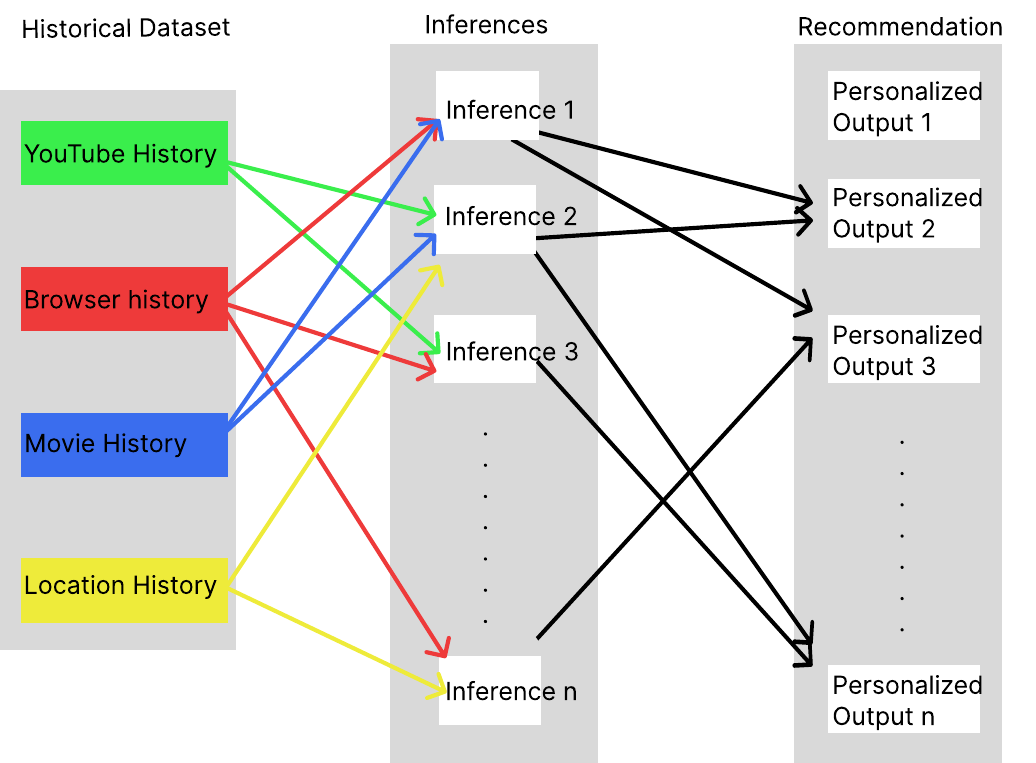}
    \caption{Sankey Visualization Prototype}
    \label{fig:prototype}
\end{figure}

Our educational artifact models the online user profiling and personalization process via a simplified but illustrative LLM-based simulation and an interactive three-layered Sankey diagram (see figure~\ref{fig:prototype}). This approach serves two purposes: (i) the simplified presentation accommodates young adults who have yet to acquire advanced Computer Science and Mathematics prerequisites~\cite{grover2023AIED}; (ii) it helps young adults effectively trace the complex relationship between digital activity, user profiling, and personalized recommendations, thereby facilitating a conceptual understanding of the personalization process. Below, we discuss the steps involved in producing the Sankey diagram. Given the sensitive nature of online activities and personal recommendations, we have deliberately implemented privacy-preserving measures, including secure user data upload facilities and a locally hosted LLM to produce the inferences, personalized recommendations and visualizations. This ensures that the data is stored on, and never leaves, a secure server approved by our IRB protocol, and that any data processing does not require us to view, open, or manually inspect any user data files, or share them with third parties.

\subsection{Step 1: Ingesting Users' Online Activities}
Users anonymously upload at least four digital activity archives to a Box folder\footnote{The Box folder is IRB-approved, institutionally-provided, does not require authentication or account creation, and does not collect uploader-side metadata.}, such as their browser history (e.g., Chrome, Safari, Mozilla), social media activity (e.g., YouTube, Instagram or Tiktok), location history (e.g., Google or Apple Maps), Search History, News history (e.g., Google News), shopping history (e.g., Amazon, Google Play Store, App Store), or entertainment history (e.g., Netflix, Hulu, Prime Video, Kindle Unlimited, Spotify)\footnote{As an example of such data, Google Takeout provides access to user data archives that comprise various services, including Chrome browsing history, YouTube, Google search, location, news, shopping, and entertainment.}. Anonymity breaks the structural link between participants, their data archives, and researchers/developers, preventing post hoc linkage between datasets and participants and thereby supporting ethical data-handling practices. After users upload their data, it is automatically processed into normalized records that contain an item name (e.g. search query, product name, video title, location), hyperlink, time, date, and the action completed on the item (e.g. ``play'', ``search'', or ``bookmark'').

\subsection{Step 2: Generating User Profiles}
Step 2 submits the user's data to a user profiling process~\cite{wang2025lettingo, middleton2004ontological, smith2023privacy} to infer what user is interested in, and what products they can be recommended/advertised accordingly. We generated the inferences and recommendations using the open-source \texttt{gpt-oss-20b} model, primarily motivated by privacy considerations. Because the input data consists of sensitive behavioral traces derived from personal digital activity, transmitting this information to third-party LLM servers would introduce unnecessary privacy risks. By keeping all computation local, the system ensures that participant data never leaves the secure processing environment. Open-source models also support reproducibility and auditability. Because both the model weights and inference scripts are publicly accessible, future researchers can replicate the entire pipeline, validate results, and modify components without dependency on closed proprietary systems. In addition, local inference ensures predictable cost and eliminates the variability associated with usage-based API pricing and rate limits. 

\subsubsection{Inferring Interests and Recommendations}
We use a single, structured prompt that jointly elicits inferred interests and personalized recommendations. The inferences and recommendations generation follows the following logic: generate \(n\) inferences for (1) each of the four datasets (total of \(n*4\) inferences), (2) each of the six datasets created by combining two datasets (total of \(n*6\) inferences), (3) each of the four datasets created by combining three datasets (total of \(n*4\) inferences), and (4) combination of all four datasets (total of \(n*1\) inferences), where we can choose a value between \(3 <= n <= 9\) to vary the complexity of the diagram. We have found that anything over nine results in theoretical saturation~\cite{saunders2018saturation}, meaning the model starts producing duplicate inferences when more than nine are requested. We prompt the model to generate a single recommendation per inference. Next, to simplify the Sankey diagram, we use the sentence-embedding model \texttt{all-MiniLM-L6-v2} to merge the same inference presented in different formats (e.g., \texttt{``interested in brain computer interfaces''} and \texttt{``interested in brain-interfacing computers''}) into a single interest. Finally, we categorize the inferences by applying agglomerative clustering with a fixed distance threshold to the cosine similarities between inference embeddings, and assigning the inference with the highest average cosine similarity to all other entries in the cluster as the category label. The model outputs a list of inferred interests, with for each inference a corresponding product recommendation, a category label, a 1--10 ``sensitivity score'' and a 1--10 ``commonness score''. The sensitivity score refers to the degree to which an inferred interest and the consequent recommendation would be telling of one's personal life, for example, a health condition. The commonness score refers to the degree to which a given person's inferred interest and the consequent recommendations are aligned with those of others in their demographic. The education artifact allows users to filter the Sankey diagram by category, sensitivity score and/or commonness score, as a means to hone in on a specific set of sensitive or unusual inferences.

\subsubsection{Mapping LLM-generated Recommendations to Amazon Products}
We use a Retrieval-Augmented Generation (RAG) component built on the Amazon Reviews 2023 dataset~\cite{amazon_reviews_2023} to associate each of the generic product recommendations produced by the LLM with a concrete example of a concrete, existing Amazon product. All products in the Amazon dataset that have both title and description are embedded using the \texttt{sentence-transformers/all-MiniLM-L6-v2} model (selected due to its balance between embedding quality and computational efficiency). The resulting embeddings are then indexed using the FAISS similarity search framework~\cite{douze2025faiss}, enabling efficient nearest-neighbor queries over the full product corpus during retrieval. To extract products relevant to the LLM-generated recommendations, the embedding of each recommendation is passed to the FAISS index, which performs a vector similarity search to identify the most semantically similar product entries. In the current implementation, the top-1 match (k=1) is retrieved and attached to the original recommendation as its grounded product example. 

\subsection{Step 3: Producing the Sankey Diagram}
We opted for a Sankey diagram~\cite{riehmann2005interactive} because it can show how different combinations of data sources can yield different inferred interests and personalized recommendations. In our three-layered Sankey diagram, the first layer represents the users' online activities, divided by source, the second layer contains the inferences, and the third layer presents the personalized recommendations. The connections between layers quantify the prevalence of the connections between specific data sources, inferences, and recommendations, and enable users to trace how their behavioral data are used to infer interests, ultimately resulting in concrete personalized recommendations. Starting from this rather complex overview, we allow users to filter the diagram by data source, inference category, inference sensitivity and commonness.
The \textit{source filter} lets users display only inferences and recommendations that originate from the selected data sources (e.g., Search, YouTube). For an inference to be visible, all sources contributing to that inference must be turned on. Any recommendations linked exclusively to hidden inferences are also hidden. This approach ensures the visualization consistently reflects how online entities with access to a certain subset of sources may make inferences and recommendations. Users can also filter by \textit{inference category}. Selecting a category restricts the display to only those inferences associated with that thematic group. Recommendations linked exclusively to hidden inferences are also hidden, and if a certain source is not involved in a particular inference category, it is hidden as well. This mode is particularly useful for users to inspect sensitive inference domains such as finance or health. A similar goal can be achieved by filtering the visualization by inference \textit{sensitivity score} and/or \textit{commonness score}. Together, these interactive mechanisms allow the visualization to express complex, multistage relationships between user behavior, LLM-generated inferences, and retrieval-grounded product recommendations in a compact and interpretable format.

%% file: 03_thinkAloudStudy.tex
\section{Educational Artifact Evaluation using a Think-Aloud study}

\begin{table}
    \centering
    \begin{tabular}{p{0.2\textwidth}p{0.65\textwidth}}
    \toprule
        Themes & Categories \\
        \midrule
         Understanding AI Processes & (i) Heightened awareness of privacy implications (\textit{``Risks of behavioral data collection feel more concrete} -- P2)\\
         & (ii) Better understanding of online activities come together to form seemingly unrelated recommendations \\
         & (iii) Ordinary behavior could be revealing of otherwise personal traits \\
         \midrule
         UI Improvements & (i) Checkbox in dropdown filters to improve usability; \\
         & (ii) Tooltip explaining the connections, especially the seemingly unrelated ones; \\
         & (iii) Filter titles should be self-explanatory \\
         \midrule
         Need for control over data & (i) Feeling unmotivated to engage in privacy-preserving behavior (\textit{``Practically speaking, I don't expect change my online behavior''} -- P1)\\
         & (ii) Practically impossible to avoid digital systems yet no control over their data \\
         & (iii) Control over data could motivate privacy preserving behavior \\
         \bottomrule
    \end{tabular}
    \caption{Table of Themes and Categories}
    \label{tab:themes_categories}
\end{table}

We conducted a pilot think-aloud study with six young adult participants (see Table~\ref {tab:dem} for demographics) upon Institutional Review Board (IRB) approval. We recruited participants via word of mouth and conducted one-hour think-aloud sessions via Zoom, which were recorded and transcribed. The goal of this pilot study was to examine whether our educational artifact helps users comprehend the relationship between digital activity and personalized recommendations. To achieve this goal without having participants upload their own data, we conducted the think-aloud studies with a visualization produced based on the first author's own personal data. All participants evaluated the same visualization, ensuring consistency of the evaluated tool. The think-aloud study consisted of seven tasks evaluating the potential of the visualization for understanding personalized recommendations and eliciting privacy-preserving behaviors. Participants were asked for (1) their initial impression of the educational artifact, (2) an explanation of its motivation and purpose. Next, participants used (3) the data source filter to inspect the provenance of inferences and recommendations, and (4) the sensitivity and uncommonness filters to see how users' interactions contribute to sensitive or uncommon inferences and their associated recommendations. Participants were then asked (5) how interacting with the artifact altered their perception of privacy and (6) whether this could entice them to take privacy-preserving actions. Finally, (7) we asked participants about the visualization's strengths, weaknesses, and areas for improvement to increase usability and comprehension.

\begin{table}
    \centering
    \begin{tabular}{llll}
    \toprule
         Participant ID & Gender & Age & Education level\\
         \midrule
         P1 & Male & 19 & College Freshman \\
         P2 & Male & 19 & College Freshman \\
         P3 & Female & 18 & College Freshman \\
         P4 & Female & 16 & High school 10th grade \\
         P5 & Male & 15 & High school 9th grade \\
         P6 & Male & 16 & High school 10th grade\\
         \bottomrule
    \end{tabular}
    \caption{Participant demographic}
    \label{tab:dem}
\end{table}

\section{Think-Aloud Results} \label{sec:results}

Thematic analysis~\cite{braun2019reflecting} (see table~\ref{tab:themes_categories} for themes and categories) of the think-aloud transcripts shows that the illustrative simulation presented as a Sankey diagram helped participants understand how online activities relate to user profiling and personalized recommendations (\textit{``I think it makes sense, because, travel, [like], the refillable, bottle kit, [like], it makes sense if you're gonna go travel. You can't bring, like, everything with you, it makes sense to have that.''} -- P4). Participants described the illustrative simulation as intuitive and appreciated the ease of tracing digital activities to the recommendations via inferences. Similarly, the ability to trace recommendations backward through inferences to specific sources was repeatedly described as improving interpretability and conceptual trust in the system. 

Consistent with extant literature (for example~\cite{sunTheySeeYou2021, sun2024cloud}), our participants' understanding of how digital activities shape their experiences was initially rather rudimentary (for example \textit{`I feel like maybe it's, [like], pattern recognition, Or,  maybe the people around me, their data, kind of compare it to my data, and see, like, maybe I'll like this'' -- P4}). While they recognized that personalized recommendations and their digital activities are related, their initial understanding of the underlying process was either incomplete or misguided. P6, for instance, thought data tracking happens through interaction with hashtags only (\textit{``Instagram uses tags in posts, so, if you see a post about, [like], a Nike shoe, it's gonna be, hashtag Nike shoe, hashtag post, okay? So, those kind of stuff, it's…being recorded on your data}). The artifact improved their understanding by highlighting that collected data persists in databases and facilitates user profiling and personalized recommendations (\textit{``I used to be, like, really into, building computers last year, and I stopped. Then, randomly, I started getting, like, a bunch, [of] ads and recommendations for them. So… I don't know why that was happening.'' -- P5}). Upon realizing their historical data persists, regardless of how much time has passed, participants were initially perplexed, followed by an understanding of why they would see certain recommendations.

All participants commented on how distinct activities (e.g., search and video consumption), could jointly contribute to a single, seemingly unrelated inferred interest. However, they were initially not able to articulate the process through which this happens. Upon probing, P4 and P6 loosely connected their existing mental models (\textit{``hashtags can be, like, transferred to, like, other sites that you are already logged into, like Gmail, or Facebook, or YouTube, hey, you could get ads on YouTube''} -- P6), which confirmed the need for comprehensive education on this topic. When asked whether their newly-gained knowledge would make any difference, they reported that they could see themselves exercising some caution during their online interactions. However, all participants felt constrained in their ability or willingness to do so meaningfully, as participation in large digital ecosystems is effectively unavoidable. They described relying on search engines, video platforms, and navigation tools as essential utilities, despite acknowledging discomfort with the extent of behavioral tracking involved. P2 reflected that although the system made underlying data practices more visible, ``practically speaking,'' they did not expect to meaningfully change how they searched, watched content, or used online services.

Finally, participants suggested the following improvements to reduce confusion and improve comprehension: (i) a checkbox-based list for filtering input data and inference categories would be more intuitive, (ii) an alternative title or a tooltip explaining what the sensitivity score refers to, (iii) a clear distinction between and explanation of online activities, specifically browser history and Google search history, (iv) more distinctive labels for each layer and (v) highlighting the connections with the colors of the input nodes (\textit{``They should be more defined, so it's easier to tell, and maybe a little bit more spaced out''} -- P6). Additionally, participants suggested adding two new features to increase comprehension: (i) adding a hover interaction emphasizing only the connections associated with a selected node while reducing the opacity of others, and (ii) contextual information explaining relationships, especially for inferences and recommendations from seemingly unrelated activities, as the visualization alone did not fully convey why such a connection existed.

%% file: 04_discussionConclusion.tex
\section{Discussion}
 
We present the development and pilot evaluation of an educational artifact that simulates the relationship between personal digital activity, user profiling and personalized recommendations via a Sankey diagram. Our pilot evaluation finds that an illustrative model simulating this otherwise abstract relationship helps young adults better understand the process. Our findings also reinforce extant literature on young adults' limited knowledge of this process~\cite{livingstone2019audiences, wang2023datafication, gao2025childrenLiteracy}, further emphasizing the necessity for such an educational artifact. While our artifact elicits awareness of how distinct activities, such as search history and location history, can jointly contribute to a single, seemingly unrelated inferred interest~\cite{Kahila2024DataTraces}, participants still lacked a clear understanding of the concept of cross-platform tracking, even after substantial prompting~\cite{sun2024cloud}. Further educational scaffolding could help users understand how algorithms employed by data aggregators integrate their online interaction as a unified unit, potentially leading to unexpected inferences. We also found that some participants initially had the misguided understanding that their data is tracked and collected \emph{only} through hashtags. By simulating users' comprehensive digital activities, we were able to partially dispel such misunderstandings.

Even after just a single interaction with the illustrative model, participants were able to reflect on how to exercise more caution and awareness in their online interactions. This finding is promising of the artifact's ability to elicit nuanced privacy reasoning, particularly among younger adults, who are known to struggle with understanding the impact of abstract AI processes~\cite{zhao2019sillyname}. However, most users followed this reflection with a realization that it is difficult to abandon the online services that track, collect, and use their data, which, for some participants, led to feelings of helplessness. This helplessness, in turn, may contribute to a defeatist attitude about adopting privacy-preserving practices online. This attitude could be indicative of a broader need for efforts to provide users with autonomy and agency~\cite{khan2025inclusive, blinder2025childrenprivacyagency, khan2024evaluating, 10.1145/3769427} over their data, and by extension, encourage the adoption of more privacy-preserving online behaviors~\cite{smith2023privacy, zhao2019make, zhao2022koala, wang2023notanumberindatabase}.

Our study is not without limitations. Although our sample size matches the median for participatory methods~\cite{caine2016samplesize}, we acknowledge that a larger sample could further solidify the effectiveness of our artifact. Secondly, our artifact is designed for saliency rather than accuracy, as our priority was young adults' comprehension. Despite our best efforts to conceptually align the inference and recommendation process with real-world practices, our simplified approach could be misleading in certain instances. Furthermore, participants evaluated a visualization of another person's data rather than their own. This lack of personal relevance means that the study likely uncovered the lower bound of the educational value of the artifact. Finally, the artifact may have a more sustained impact when embedded in a classroom module providing scaffolding instructions and explanations.

\section{Conclusion}
This paper presents the development and preliminary evaluation of an educational artifact that illustratively simulates how one's digital activities shape their experiences online. While the illustrative model effectively promotes comprehension and reflection about privacy implications, we find that users may not necessarily change their online behavior, as they do not see an alternative to using digital platforms and means of protecting their data. 

\section{Declaration on Generative AI}
During the preparation of this work, the author(s) used Grammarly in order to perform grammar and spelling checks. After using this tool/service, the author(s) reviewed and edited the content as needed and take(s) full responsibility for the publication’s content.